\documentclass[aps,pra,twocolumn,showpacs,superscriptaddress,groupedaddress]{revtex4-1}
\usepackage[margin={2cm,2cm}]{geometry}
\usepackage{graphicx}
\usepackage{natbib}
\usepackage{amsthm}
\usepackage[english]{babel}

\usepackage{amsmath}
\usepackage{amsfonts}
\usepackage{amssymb}
\usepackage{subfig}
\usepackage{hyperref}
\usepackage[usenames, dvipsnames]{color}
\usepackage{bbold}
\usepackage{bbm}
\usepackage{amsfonts}
\usepackage{hyperref}
\usepackage[english]{babel}
\hypersetup{
    colorlinks=true,
    linkcolor=ForestGreen,
    filecolor=magenta,      
    urlcolor=cyan,
    citecolor=blue
}
\usepackage{epstopdf}
\usepackage{lipsum}

\begin{document}
 
 \title{Entanglement assisted training algorithm for supervised quantum classifiers}
  \author{Soumik Adhikary}
  \email{soumikadhikary@physics.iitd.ac.in}
  \affiliation{Department of Physics, Indian Institute of Technology Delhi, New Delhi-110016, India.}
  
\begin{abstract}  We propose a new training algorithm for supervised quantum classifiers. Here, we have harnessed the property of quantum entanglement to build a model that can simultaneously manipulate multiple training samples along with their labels. Subsequently a Bell-inequality based cost function is constructed, that can encode errors from multiple samples, simultaneously, in a way that is not possible by any classical means. We show that upon minimizing this cost function one can achieve successful classification in benchmark datasets. The results presented in this paper are for binary classification problems. Nevertheless, the analysis can be extended to multi-class classification problems as well.
\end{abstract}
  
  %\pacs{03.67.Ac, 03.67.Lx, 03.65.Ud}

\maketitle

\section{Introduction}

Machine learning (ML) has emerged as an important area of research in recent years. The success of ML may be attributed to its wide range of applications, such as in image recognition, drug discovery, finance, material design, etc. \cite{autoencoders2016, lecun2015, medical2018, guzik2018}. Quantum computation, on the other hand, is a fundamentally new way of computing based on the principles of quantum mechanics. It has been shown that there can be a number of advantages that quantum computation may offer over their classical counterparts \cite{Shor1994, Grover1997, QKD2009, bosonsampling2013, bosonsampling2017}. Following this observation, it has been realised that ML algorithms too could gain from the use of quantum computing \cite{biamonte2017_review, schuld2014, farhi2018, benedetti2019, beer2020, blank2020}.  Subsequently, several quantum machine learning (QML) algorithms were developed, that has conclusively shown that quantum computing can indeed assist ML algorithms, in the form of speedups \cite{wossing2018}, estimating classically intractable kernels, etc. \cite{havlivcek2019, schuld2019}

Of particular interest to us are the QML algorithms involving the so-called variational quantum circuits \cite{mcclean2016, havlivcek2019, CQC2019, maria2020}. These are  hybrid quantum-classical models that can be implemented on noisy intermediate-scale quantum (NISQ) devices with relative ease. A variational quantum circuit comprises of a parametrised ansatz; the parameters are trained to accomplish the desired QML task. Typically it is only during the forward pass of the algorithm where quantum effects are explicitly used. The forward pass in a variational quantum circuit comprises of three stages - state preparation, operation of a parametrised ansatz and, finally, a measurement operation \cite{mitarai2018, maria2020}. The data collected from the measurement operation is then classically post-processed to get the cost function, which in turn is minimised iteratively to train the circuit (more particularly the parametrised ansatz) \cite{harrow2019, schuld2019opt}. Clearly, the cost function is a classical quantity. The details of the classical post-processing depend on the nature of the problem. For example, in a classification task under supervised learning, the information on the training samples' labels is used to post-process the outcomes of the measurement operation. The resultant cost function is typically a cross-entropy error or simply the norm of the difference between the expected measurement outcome (determined by  the label) and the obtained measurement outcome. The process of calculating the cost function is very similar to how they are calculated for classical ML models, such as in neural networks. The training samples are processed {\it individually} through the classifier model (in this case, a quantum circuit). The output  of the model (in this case, the measurement outcome) for each sample and the corresponding labels are then used to calculate the cost function. Recently, attempts have been made to embed the cost functions in the quantum circuit directly so that they can be manipulated quantum mechanically \cite{cao2020}.

The purpose of this paper is to propose a new training algorithm for variational quantum circuits under the supervised learning scheme. The proposed model is for binary classification, although it can be generalised to multi-class classification problems as well. We show that by encoding the label of a training sample directly into the variational quantum circuit, along with the sample itself, one can harness genuine quantum effects such as entanglement to construct novel cost functions. Such cost functions do not have any classical counterpart and are therefore unique to QML models. We show that it is possible to embed the information for multiple training samples into such cost functions, simultaneously. The precise context in which the term simultaneous is used here will become clear later in the paper. However it must be emphasized that the proposed method is fundamentally different from its classical counterpart in the sense that that it does not involve  accumulation of individual errors before finally updating the cost function for multiple samples. In fact we will show that it is impossible to determine the errors from the individual samples, in our method. As noted earlier, most present day QML algorithms employ the classical routine for error calculation. Our method provides an alternate to this approach that is uniquely quantum mechanical.

The paper is organised as follows: In section-\ref{vqc} we present a brief introduction to supervised learning in variational quantum circuits and how they are conventionally trained. Special importance is given to dressed quantum circuits (a variant of variational circuits) as they are central to the present study \cite{adhikary2020, mari2019}.   Section-\ref{training_bell} contains the central result of this paper - an entanglement assisted algorithm for training variational quantum circuits. In the following section, we present a few numerical examples demonstrating successful data classification using our model. We discuss the results and the associated subtleties and finally conclude the paper in section-\ref{disc}.

\section{Preliminaries : supervised learning in variational quantum circuits}
\label{vqc}

\begin{figure}
\centering
\includegraphics[width=0.9\linewidth]{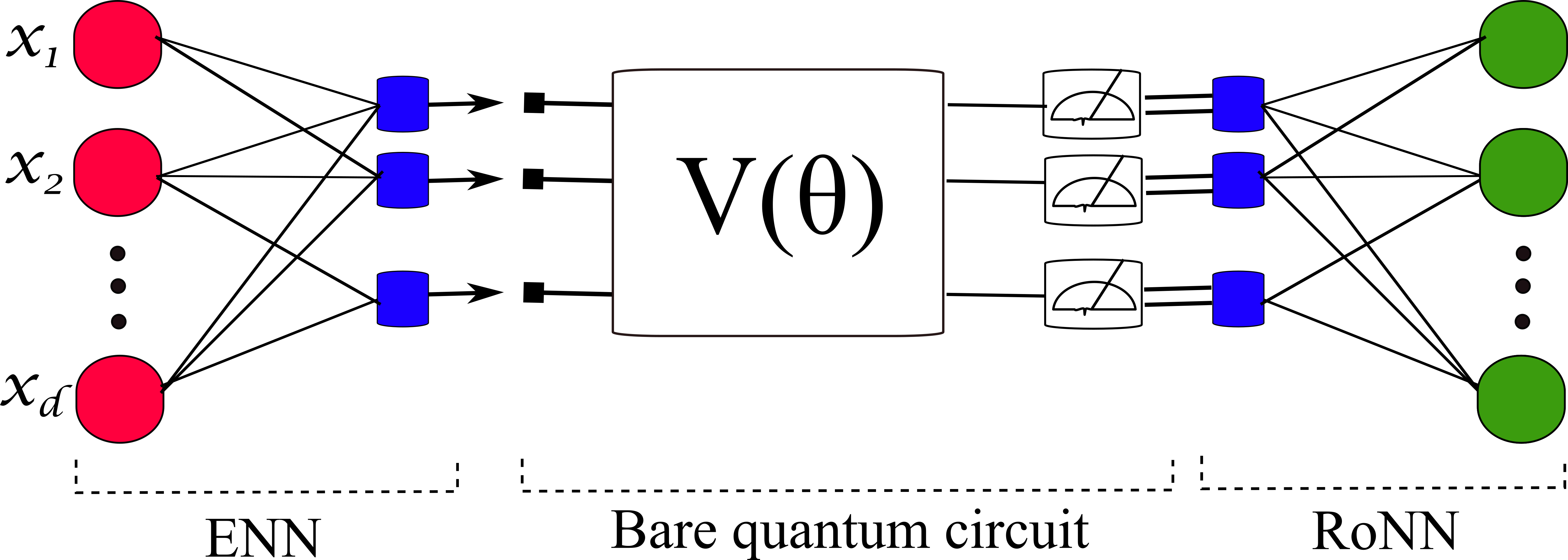}

\caption{\label{DQC} Schematic of a dressed quantum circuit}
\end{figure}

Variational quantum circuits, as mentioned before, are hybrid quantum-classical models used for various QML applications. The application that we are interested in, for the purpose of this paper, is data classification using supervised learning \cite{bishop2006, haykin2010}.  Consider a dataset ${\cal S} = \{ {\bf x}_i, f({\bf x}_i) \}_{i = 1}^N$. Each entry in ${\cal S}$ is an ordered pair of a sample ${\bf x}_i \in \mathbb{R}^d$ and an associated label $f({\bf x}_i) \in \{+,-\}$. The purpose of supervised learning is to train a parametrised model (in this case a variational quantum circuit), based on the labelled data, randomly sampled from the train set ${\cal T} \subset {\cal S}$. At the end of the training process, we expect the model to infer correct labels for all unlabelled sample ${\bf x}$ chosen from the dataset ${\cal S}$. Mathematically speaking, the model is trained to return a function $f'({\bf x}, \boldsymbol{\theta}^*)$ which must ideally be equal to $f({\bf x})$; ${\boldsymbol{\theta}}$ are the model parameters, while ${\boldsymbol{ \theta}}^*$ are the optimal (or trained) model parameters. The relation $f'({\bf x}, {\boldsymbol{ \theta}}^*) \sim f({\bf x})$ is ensured by minimizing the cost function ${\cal C} (\boldsymbol{\theta}) = g (\vert f'({\bf x}, {\boldsymbol{ \theta}}) - f({\bf x}) \vert)$, with respect to the parameters ${\boldsymbol{ \theta}}$. The cost function is minimised iteratively, typically via gradient based optimization methods. Some gradient free methods have also been explored \cite{gradfree2019}. Note that, initially the parametrised model (variational circuit) generates the function $f'({\bf x}, {\boldsymbol{ \theta}})$, independent of any information on $f({\bf x})$. It is only in the subsequent iterations of cost minimization, that $f'({\bf x}, {\boldsymbol{ \theta}})$ gets modified based on the information of $f({\bf x})$ that is contained in the cost function ${\cal C} (\boldsymbol{\theta})$, thus leading to $f'({\bf x}, {\boldsymbol{ \theta}}) \rightarrow f'({\bf x}, {\boldsymbol{ \theta}}^*)$, over a number of iterations. The variational circuit itself does not contain any information of $f({\bf x})$ explicitly. 

The process of generating $f'({\bf x}, {\boldsymbol{ \theta}})$ on a variational quantum circuit can broadly be divided into three steps. The first step is data embedding and state preparation ${\bf x} \rightarrow \vert \psi ({\bf x}) \big>$. $\vert \psi ({\bf x}) \big>$ is typically a multi-qubit state. Commonly used encoding techniques are qubit encoding, amplitude encoding, basis encoding, etc. \cite{wiebe2012, stoudenmire2016, farhi2018}. The second step involves building a parametrised quantum circuit, otherwise know as an ansatz $V({\boldsymbol{ \theta}})$, that operates on the state $\vert \psi ({\bf x}) \big>$; $\vert \psi ({\bf x}, {\boldsymbol{ \theta}}) \big> = V({\boldsymbol{ \theta}}) \vert \psi ({\bf x}) \big>$. The ansatz can vary a lot, depending on the nature of the problem. In the final step, the state $\vert \psi ({\bf x}, {\boldsymbol{ \theta}}) \big>$ is measured, usually in the eigenbasis of the third Pauli matrices - $\sigma_3 \otimes \sigma_3 \otimes \cdots$. The measurement outcome is then classically post-processed to obtain $f'({\bf x}, {\boldsymbol{ \theta}})$.

The so-called dressed quantum circuits are a special type of variational quantum circuits \cite{adhikary2020, salinas2020, mari2019}. Even though they both operate on the same working principles, the former manages to get rid of various practical drawbacks that the latter encounters.  A dressed quantum circuit typically contains a variational quantum circuit (often referred to as a bare quantum circuit in this context) with classical neural networks attached at both ends (see fig.-\ref{DQC}). One of the neural networks embeds data into the bare circuit while the other reads data out from the circuit. We call them embedding neural networks (ENN) and readout neural networks (RoNN), respectively. ENN play an instrumental role in dimensionality reduction of the original data; ${\cal N}_{d \rightarrow d'} ({\bf x}) = {\bf x'}$, ${\bf x} \in \mathbb{R}^d, {\bf x}' \in \mathbb{R}^{d'}, d' \leq d$. This facilitates the data encoding process since a lower-dimensional data can be encoded into a smaller quantum system (fewer qubits). They can also be used for feature extraction so that the bare quantum circuit has to process only a few highly informative features \cite{mari2019}. This reduces the overall cost of computation on the quantum hardware; NISQ devices suffer from several limitations, which restricts us from executing very complicated computations on them \cite{Preskill2018, ibmq}.  In a dressed quantum circuit, the parameters in the ENN and the RoNN are also trained in addition to the parameters in the bare quantum circuit.

\subsection{Training a variational quantum circuit}

To train a circuit, we start by dividing the dataset into a train set ${\cal T} \subset {\cal S}$ and a test set ${\cal S} - {\cal T}$. The train set is randomly sampled form ${\cal S}$, which is used to construct the cost function ${\cal C} (\boldsymbol{\theta})$ which is subsequently minimised:

\begin{equation}
{\cal C} (\boldsymbol{\theta}) = \frac{1}{\vert {\cal T} \vert} \sum_{i = 1}^{\vert {\cal T} \vert} g( \vert f'({\bf x}_i, {\boldsymbol{ \theta}}) - f({\bf x}_i) \vert); \ \ {\bf x}_i \in {\cal T}
\end{equation}

For a binary classification problem the function reduces to:

\begin{eqnarray}
\label{eq:bin_cost}
{\cal C} (\boldsymbol{\theta}) = \frac{1}{\vert {\cal T} \vert}  \sum_{i = 1}^{\vert {\cal T^+} \vert} &g&( \vert f'({\bf x}^+_i, {\boldsymbol{ \theta}}) - 1 \vert) \nonumber \\
&+& \frac{1}{\vert {\cal T} \vert}  \sum_{i = 1}^{\vert {\cal T^-} \vert} g( \vert f'({\bf x}^-_i, {\boldsymbol{ \theta}}) + 1 \vert)
\end{eqnarray}

Here, ${\cal T}^{\pm}$ are sets of training samples corresponding to the labels $f({\bf x}) = \pm 1$; ${\cal T} = {\cal T}^+ \cup {\cal T}^-$, ${\bf x}^{\pm} \in {\cal T}^{\pm}$. A popular choice for $g(\cdot)$ is simply $g(x) = x$. Conventionally, the training samples ${\bf x}_i$ are processed individually through the circuit to obtain the output function $f'({\bf x}_i, {\boldsymbol{ \theta}})$. The cost function ${\cal C} (\boldsymbol{\theta})$ is subsequently evaluated in accordance to Eq.~\ref{eq:bin_cost}, by collecting $f'({\bf x}_i, {\boldsymbol{ \theta}})$ for all training samples. Alternately, one can also train the circuit by taking a classical mixture of the states $\{ \vert \psi ({\bf x}_i) \big>\}$ ; ${\bf x}_i \in {\cal T}$ \cite{cao2020}. Such a system is represented by a mixed state $\rho = \frac{1}{\vert {\cal T} \vert} \sum_i \vert \psi ({\bf x}_i) \big> \big< \psi ({\bf x}_i) \vert$. The training would involve a single sample being processed for every run of the algorithm. The final cost is the average of the costs over all runs. These cost evaluation routines are clearly akin to standard methods of cost evaluation in classical ML models such as in classical neural networks \cite{haykin2010}. 

\begin{figure*}
  
  \includegraphics[width = \textwidth]{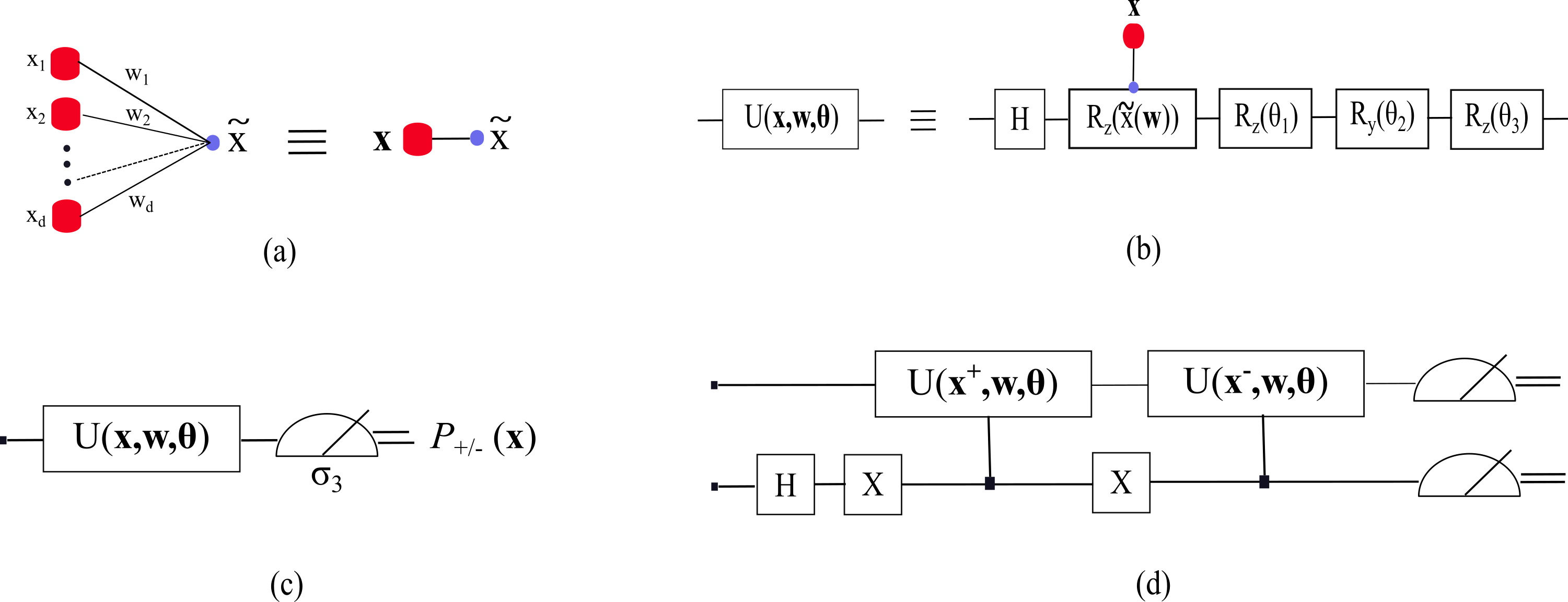}
  \caption{\label{circuit} A schematic showing (a) the ENN (${\cal N}_{d \rightarrow 1}$) used in our dressed quantum circuit (classifier). The ENN performs a vector matrix multiplication operation that takes  a $d$ dimensional input vector (sample) ${\bf x}$ to a single number ${\tilde{x}} = \sum_i x_i w_i$. (b) The gate operations used in our classifier circuit, to encode a sample ${\bf x}$ into a quantum state, followed by a parametrised SU(2) operation (Eqs.~\ref{eq:state_prep_class} and \ref{eq:state_rot_class}) (c) The dressed quantum circuit: our classifier model. (d) The training circuit: the circuit can process two training samples at a time (${\bf x^+}, {\bf x^-}$). The circuit is used to find the optimal parameters in our classifier model (shown in (c)), for efficient data classification. The qubit at the top is the sample qubit while the one at the bottom is the label qubit. In both (c) and (d) the qubits were initialised at its ground state.}
\end{figure*}

\section{Result : an entanglement assisted training algorithm}
\label{training_bell}

This section presents the main result of this paper. We propose a new method to  train variational quantum circuits (in this case, a dressed quantum circuit) within the framework of supervised learning. Our method is fundamentally different from how such circuits are conventionally trained. More specifically, there are two key ways in which our method differs from the conventional ones. Firstly we encode the label along with the corresponding training sample, directly into the variational circuit, during state preparation. This allows us to manipulate  the labels, quantum mechanically. Recall that the predominant practice is to process the information on the label only while calculating the cost function, which is entirely a classical process. Secondly and more importantly, we use entangled states to encode multiple training samples along with their labels, in order to train our variational circuit. We harness the non-classical correlations in these states to yield a Bell-inequality based cost function. We show that this cost function captures the errors for all the training samples that are encoded in the entangled state, without having to evaluate the error contribution from the individual samples explicitly.  In-fact it is impossible to determine the individual errors in our method. Thus the manner in which our Bell-inequality based cost function is evaluated is essentially quantum mechanical, with no classical counterpart. It inherits this unique property from their parent entangled states, where the encoded samples can not be treated as separate entities. In this sense we can claim that our cost function can encode the errors for multiple training samples - simultaneously.

\subsection{The dressed quantum circuit}
\label{DQC_expl}

We shall use the dressed quantum circuit that was originally proposed in \cite{adhikary2020} as our classifier model. It consists of a simple ENN, which is a fully connected classical neural network with $d$ input nodes, a single output node, no hidden layers in between, and no activation functions. The ENN is symbolically represented as ${\cal N}_{d \rightarrow 1}$. The purpose of this layer is to perform a vector matrix multiplication operation that takes a $d$ dimensional input vector (sample) to a single number; ${\cal N}_{d \rightarrow 1} ({\bf x}) = \tilde{x}({\bf w}) = \sum_i w_i x_i$ (see fig.-\ref{circuit}(a)). ${\bf w}$, collectively, are the weights of the neural network. The ENN is used just for dimensionality reduction. It does not perform any other advanced operation, such as feature extraction. Hence the entire classification takes place on the bare quantum circuit.

The classically pre-processed vector is then fed into a bare quantum circuit with just a single qubit (for binary classification). The vector is first encoded into a quantum state:

\begin{equation}
\label{eq:state_prep_class}
\vert \psi ({\bf x}, {\bf w}) \big> = e^{i \sigma_3 \tilde{x} ({\bf w})} H \ \vert 0 \big>
\end{equation}
where $H$ is the hadamard gate and $\sigma_3$ is the third Pauli matrix. Next, the state is allowed to undergo an arbitrary rotation (a parametrised $SU(2)$ operation):

\begin{equation}
\label{eq:state_rot_class}
\vert \psi ({\bf x}, {\bf w}, \boldsymbol{\theta}) \big> = e^{i \sigma_3 \theta_3} e^{i \sigma_2 \theta_2} e^{i \sigma_3 \theta_1} \vert \psi ({\bf x}, {\bf w}) \big>
\end{equation}
and finally, it is measured in the eigenbasis of $\sigma_3$ (see fig.-\ref{circuit}(b)(c)). Our dressed circuit does not contain a RoNN. The outcome of the projective measurement, therefore, is directly used to classify data. The metric chosen is simple. The label $f({\bf x}) = +$ is assigned to a sample ${\bf x}$ if the probability of the outcome $\sigma_3 = -1$, for the state $\vert \psi ({\bf x}, {\bf w}, \boldsymbol{\theta}) \big>$, is greater than 0.5; $P_- ({\bf x}) \geq 0.5$. On the other hand, the label $f(\bf x) = -$ is assigned to a sample if the probability of the outcome $\sigma_3 = +1$, for the corresponding quantum state, is larger than 0.5; $P_+ ({\bf x}) \geq 0.5$. Thus for correct classification, we would require $P_- ({\bf x}^+) \geq 0.5$ and $P_+ ({\bf x}^-) \geq 0.5$. However, ideally, we would want the circuit to be trained such that $P_- ({\bf x}^+) \rightarrow 1$ and $P_+ ({\bf x}^-) \rightarrow 1$. Physically  this would mean that under ideal conditions we expect the mappings:

\begin{equation}
\label{mappings:eq}
 \vert \psi ({\bf x}^+, {\bf w}, {\boldsymbol{ \theta}}) \big> \rightarrow \vert 1 \big> , \ \ \  \vert \psi ({\bf x}^-, {\bf w}, {\boldsymbol{ \theta}}) \big> \rightarrow \vert 0 \big>. 
\end{equation}
The conditions in Eq.~\ref{mappings:eq} can be achieved by minimizing a suitably chosen cost function ${\cal C} ({\bf w}, {\boldsymbol{\theta}})$, with respect to the parameters ${\bf w}$ and ${\boldsymbol{\theta}}$ such that ${\cal C} ({\bf w}^*, {\boldsymbol{\theta}}^*) \sim 0$ ; ${\bf w}^*, {\boldsymbol{\theta}}^*$ are the optimal (or trained) parameters. The principle objective of training a circuit is to identify these optimal parameters ; they lead to maximum classification accuracies.

\subsection{Training}

Consider a train set ${\cal T}$ randomly sampled from ${\cal S}$. We assume that ${\cal T}$ contains an equal number of samples from both classes; $\vert {\cal T}^+ \vert = \vert {\cal T}^- \vert = m$, ${\cal T} = {\cal T}^+ \cup {\cal T}^-$, ${\bf x}^{\pm} \in {\cal T}^{\pm}$. To train the dressed circuit described earlier, we first stipulate under our new training scheme, that the label of the training data must also be encoded into a quantum state, along with the sample itself. A simple way to do this is via the mapping ${\bf x} \rightarrow \vert \psi ({\bf x}, {\bf w}, {\boldsymbol{\theta}}) \big> \otimes \vert L \big>$. $\vert L \big>$ is the state of an extra qubit that takes care of the label. We call the first qubit, the one in which a sample is encoded, as the sample qubit. The second qubit, that contains the information of the label, is the label qubit. $L = 0$ when $f ({\bf x}) = +$ and $L = 1$ when $f ({\bf x}) = -$. Next, we arrange the elements in the training set ${\cal T}$ as ordered pairs of form $({\bf x}^+, {\bf x}^-)$ (in no particular order); ${\cal T} = \{({\bf x}^+_i, {\bf x}^-_i) \}_{i = 1}^{m}$. As per our assumption, the training set contains an equal number of samples belonging to both the classes. Hence it is always possible to arrange the train samples as prescribed. We now encode these pairs into the states:
\begin{equation}
({\bf x}_i^+, {\bf x}_i^-) \rightarrow \frac{1}{\sqrt{2}} (\vert \psi ({\bf x}^+_i, {\bf w}, {\boldsymbol {\theta}}) \big> \vert 0 \big> + \vert \psi ({\bf x}^-_i, {\bf w}, {\boldsymbol {\theta}}) \big> \vert 1 \big>)
\end{equation}
Henceforth, for the sake of brevity, we shall denote these states as $\vert \Psi^i \big>_L$. $\vert \Psi^i \big>_L$ allows us to encode pairs of labelled training samples belonging to different classes, along with their respective labels, into a single entangled state. The entangling circuit shown in fig.-\ref{circuit}(d) can be used to prepare 
$\vert \Psi^i \big>_L$. We call this - the training circuit. It is clearly distinct from the dressed quantum circuit, which we have chosen as our classifier model (fig.-\ref{circuit}(c)). Nevertheless, we show that the training circuit can help  us  identify the  optimal parameters - ${\bf w}^*$ and ${\boldsymbol{\theta}}^*$ - 
for our classifier model , thus {\it training} the dressed quantum circuit. Introducing a separate training circuit,  allows us to develop novel cost functions, that can explicitly harness the entanglement properties of the state $\vert \Psi^i \big>_L$. To accomplish this, we perform a Bell test \cite{CHSH1969, brunner2014} on the state $\vert \Psi \big>_L$. It is well known that the Bell inequality gets violated for all entangled pure two-qubits states \cite{gisin1991}. The inequality gets maximally violated if the state is maximally entangled. More mathematically, the expectation of the Bell operator ${\cal B}$ attains the Tsirelson bound \cite{cirel1980} for maximally entangled states; $\vert \big< {\cal B} \big> \vert = 2\sqrt{2}$. Obviously, this happens for specific configurations of observables in ${\cal B}$. Consider the following configuration for an example \cite{braunstein1992}. Let there be a pair of orthogonal dichotomic observables - $\sigma_1$ and $\sigma_2$ for the sample qubit and another pair of orthogonal dichotomic observables $(\sigma_1 + \sigma_2)/\sqrt{2}$ and $(\sigma_1 - \sigma_2)/\sqrt{2}$ for the label qubit. $\sigma_1$ and $\sigma_2$ are the Pauli operators. The resultant Bell operator for this configuration becomes:

\begin{equation}
{\cal B} = \sqrt{2} (\sigma_1 \otimes \sigma_1 + \sigma_2 \otimes \sigma_2)
\end{equation}
It is easy to see that $\vert \big< {\cal B} \big> \vert = 2\sqrt{2}$ for the state $\frac{1}{\sqrt{2}} (\vert 1 0 \big> \pm \vert 0 1 \big>)$, also known as the Bell state. We shall use this result to train the circuit. We formally define the cost function for each pair of training sample $({\bf x}_i^+, {\bf x}_i^-)$  as:

\begin{equation}
{\cal C}_i ({\bf w}, {\boldsymbol{\theta}}) =  2\sqrt{2} - \vert \big< {\cal B} \big>_L^i \vert
\end{equation}
where $ \big< {\cal B} \big>_L^i =  \big< \Psi^i \vert {\cal B} \vert \Psi^i \big>_L$.  The rationale for making such a choice is to ensure that the quantity  $\vert \big< {\cal B} \big>_L^i \vert$ goes to $2\sqrt{2}$ for every $\vert \Psi^i \big>_L$, upon minimizing ${\cal C}_i ({\bf w}, {\boldsymbol{\theta}})$. Physically this would imply - the mapping $\vert \Psi^i \big>_L \rightarrow \frac{1}{\sqrt{2}}(\vert 1 0 \big> \pm \vert 0 1 \big>)$ and hence $\vert \psi ({\bf x}_i^+, {\bf w}, {\boldsymbol{ \theta}}) \big> \rightarrow \vert 1 \big>$ and $\vert \psi ({\bf x}^-_i, {\bf w}, {\boldsymbol {\theta}}) \big> \rightarrow \vert 0 \big>$. The latter is the condition that we wish to achieve (ideally) when ${\bf w} \equiv {\bf w}^*$ and $\boldsymbol{\theta} \equiv {\boldsymbol{\theta}}^*$ in our classifier circuit, as discussed in Eq.~\ref{mappings:eq}. It is therefore clear that, in the current approach, successful training can be achieved, without having to evaluate the individual errors for ${\bf x}^+$ or ${\bf x}^-$; The quantity ${\cal C}_i ({\bf w}, {\boldsymbol{\theta}})$ captures the errors for the pair $({\bf x}^+, {\bf x}^-)$ simultaneously, in a manner that is entirely quantum mechanical. The total cost function is obtained by averaging over the cost function for all such pairs of training samples:

\begin{equation}
\label{tot_cost}
{\cal C} ({\bf w}, {\boldsymbol{\theta}}) =  \frac{1}{m}\sum_{i = 1}^{m} \big( 2\sqrt{2} - \vert \big< {\cal B} \big>_L^i \vert \big)
\end{equation}
It is the total cost function that is minimised to train our classifier model.

An efficient way of constructing the cost function ${\cal C} ({\bf w}, {\boldsymbol{\theta}})$ is to use a classical mixture of the states $\{ \vert \Psi^i \big>_L \}_{i = 1}^{m}$ \cite{cao2020}. Such an ensemble is represented by a density matrix:
 
\begin{equation}
\label{mixed_state_prep_et}
\rho_L = \frac{1}{m} \sum_{i = 1}^{m} \vert \Psi^i \big> \big< \Psi^i \vert_L
\end{equation}
This would entail a single pair $({\bf x}_i^+, {\bf x}_i^-)$ being processed by the training circuit to calculate the cost function ${\cal C}_i ({\bf w}, \boldsymbol{\theta})$ in each run of the algorithm. The total cost is an average of the outcomes over all runs of the algorithm. 

Instead of using the mixed state $\rho_L$ one may also choose to consider a coherent superposition of all the sample states $\{ \vert \Psi^i \big>_L \}_{i = 1}^{m}$. We follow the procedure in \cite{cao2020} to construct  such a superposition that can be used to evaluate ${\cal C} ({\bf w}, \boldsymbol{\theta})$; $\vert \Phi \big> = \frac{1}{\sqrt{m}} \sum_i \vert \Psi^i \big>_L \otimes \vert \epsilon^i \big>$. $\vert \epsilon^i \big>$ are the basis states of a multi-qubit system also called as index qubits. However such a superposition besides being hard to prepare, also does not provide any additional advantage, as it has been shown to reduce to the mixed state ($\rho_L$) encoding \cite{cao2020}.

\begin{figure*}
  
  \subfloat[]{\includegraphics[width=0.33\textwidth]{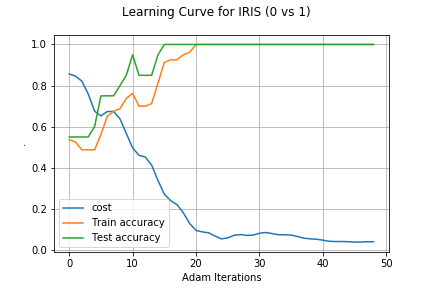}\label{01}}
  \hfill
  \subfloat[]{\includegraphics[width=0.33\textwidth]{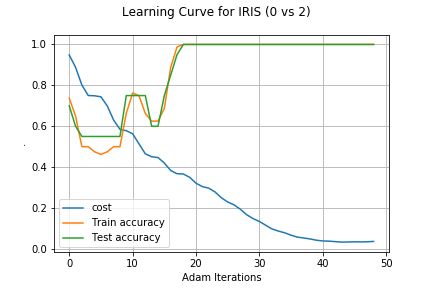}\label{02}}
  \hfill
  \subfloat[]{\includegraphics[width=0.33\textwidth]{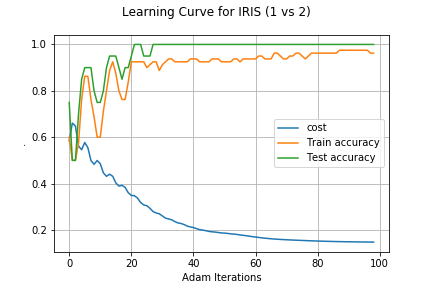}\label{12}}
  \caption{\label{fig:learning} Variation of cost, train accuracy and test accuracy with number of epochs for the three binary classification problems in the Fisher's Iris dataset-  (\ref{01}) setosa vs virginica (0 vs 1) , (\ref{02}) setosa vs versicolor (0 vs 2), and (\ref{12}) virginica vs versicolor (1 vs 2). The cost function has been suitably normalised to limit its value in $[0,1]$. Note that this does not alter the outcomes of the algorithm. The cost function was minimised over 50 iterations for (\ref{01}), (\ref{02}) and over 100 iterations for (\ref{12}).} 
\end{figure*}

\section{Numerical experiments}
\label{classification}

The effectiveness of our method is demonstrated in this section. We have considered the classification problem for the Fisher's Iris dataset \citep{FisherIris}. The dataset contains  samples belonging to the three Iris flower species - setosa (class-0), virginica (class-1) and versicolor (class-2). There are a total of 150 samples; 50 samples from each class. Each sample has four features; ${\bf x} \in \mathbb{R}^4$. The three classes are linearly inseparable. We are however, not interested in the three-class classification problem for the purpose of this paper. We will keep our analysis restricted to binary classification only. Three binary classifications are possible - class-0 versus class-1 (linearly separable), class-0 versus class-2 (linearly separable), and class-1 versus class-2 (linearly inseparable).  For each of these classification problems, we have chosen a training set of 80 samples (40  randomly selected samples per class) and a test set of 20 samples (10 randomly selected samples per class). The model was trained until convergence using the Adam optimiser \cite{adam2014}. The classification accuracies, as obtained from numerical simulations of our algorithm, are listed in Table-\ref{tab:accuracy}. The learning curves in fig.-\ref{fig:learning} shows the convergence of the cost function and the subsequent variations in the classification accuracies.

\begin{table}[]
\begin{center}
\begin{tabular}{| l | l | l |}  
\hline
%{\bf Dataset} & \multicolumn{2}{|c|}{{\bf Code }} & \multicolumn{2}{|c|}{{\bf Qiskit}} & \multicolumn{2}{|c|}{{\bf IBM-Q5}} \\  \cline{2-7}
{\bf Dataset} & {\bf Train accuracy} & {\bf Test accuracy} \\ \hline
0 vs 1 &  100 $\%$ & 100$\%$ \\ \hline
0 vs 2 &  100$\%$ &  100$\%$ \\ \hline
1 vs 2 & 96.25$\%$ & 100$\%$ \\ \hline
\end{tabular}
\end{center}
\caption{\label{tab:accuracy} Classification accuracies for the train and the test sets.}
\end{table}

\section{Conclusion}
\label{disc}

In conclusion, we have proposed a new training algorithm for binary classification in dressed quantum circuits. In particular we have introduced a novel Bell-inequality based cost function that can be used to train such models. Our cost function has no classical counterpart and is explicitly based on quantum correlations. We identify that it is key to treat the labels of the training samples, quantum mechanically, along with the sample themselves. We show that this allows us to encode pairs of training samples, with opposite labels, into entangled states. A Bell test on these states leads us to our cost function. As a direct consequence of this explicit use of non-classical correlations, we were able to show that our Bell-inequality based cost function can capture the errors corresponding to pairs of training samples (encoded in the entangled states), simultaneously, without having to explicitly calculate the errors for the individual samples.

Although the discussions in this paper are restricted to binary classification problems, our method can be extended to problems with multiple classes, as well. For example, quantum classifiers based on multi-level systems have been proposed in  \cite{adhikary2020}, for multi-class classification. Adapting our analysis to such classifiers would require us to create entangled states in coupled multi-level systems. The challenge however, would lie in selecting a suitable witness (an equivalent of the Bell inequality used here) for coupled higher dimensional Hilbert spaces \cite{collins2002, sandhir2017}, that would lead to a new cost function which would be analogous to the Bell-inequality based cost function introduced here. It may also be possible to design new cost functions following the method proposed here, that can capture the errors for more than just pairs of samples. Such an algorithm would most likely involve multiple qubits in the training circuit followed by a suitable multi-qubit entanglement witness.

\section{Acknowledgement}
The author thanks V. Ravishankar, Sooryansh Asthana, Siddharth Dangwal and Rajni Bala for fruitful discussions.


%\bibliographystyle{apsrev4-2} 
%\bibliography{sample}

\end{document}